\documentclass[%
 aip,
 amsmath,amssymb,
 reprint,%
]{revtex4-1}

\usepackage{graphicx}
\usepackage{dcolumn}
\usepackage{bm}

\usepackage[utf8]{inputenc}
\usepackage[T1]{fontenc}
\usepackage{mathptmx}
\usepackage{etoolbox}

\usepackage{subfig}
\usepackage{url}
\usepackage{dblfnote}

\graphicspath{{./pdf/}{../jpeg/}}

\makeatletter
\def\@email#1#2{%
 \endgroup
 \patchcmd{\titleblock@produce}
  {\frontmatter@RRAPformat}
  {\frontmatter@RRAPformat{\produce@RRAP{*#1\href{mailto:#2}{#2}}}\frontmatter@RRAPformat}
  {}{}
}%
\makeatother
\begin{document}

\preprint{AIP/123-QED}

\title[Surrogate-based cross-correlation for particle image velocimetry]{Surrogate-based cross-correlation for particle image velocimetry}
\author{Yong~Lee}\affiliation{School of Mechanical and Electronic Engineering, Wuhan University of Technology~(WHUT), Wuhan 430070, China
}

\author{Fuqiang~Gu}%
\affiliation{ 
College of Computer Science, Chongqing University, Chongqing 400044, China
}%

\author{Zeyu~Gong}
\affiliation{
State Key Laboratory of Intelligent Manufacturing Equipment and Technology,
School of Mechanical Science and Engineering, Huazhong University of Science and Technology~(HUST), Wuhan 430074, China}

\author{Ding~Pan}
\author{Wenhui~Zeng$^*$}%
\email{zengwenhui3242@gmail.com}
\affiliation{School of Mechanical and Electronic Engineering, Wuhan University of Technology~(WHUT), Wuhan 430070, China}


\date{\today}

\begin{abstract}

This paper presents a novel surrogate-based cross-correlation (SBCC) framework to improve the correlation performance for practical particle image velocimetry~(PIV). The basic idea is that an optimized surrogate filter/image, replacing one raw image, will produce a more accurate and robust correlation signal. Specifically, the surrogate image is encouraged to generate perfect Gaussian-shaped correlation map to tracking particles (PIV image pair) while producing zero responses to image noise (context images). And the problem is formularized with an objective function composed of surrogate loss and consistency loss. As a result, the closed-form solution provides an efficient multivariate operator that could consider other negative context images.
Compared with the state-of-the-art baseline methods (background subtraction, robust phase correlation, etc.), our SBCC method exhibits significant performance improvement (accuracy and robustness) on the synthetic dataset and several challenging experimental PIV cases. Besides, our implementation with experimental details (\url{https://github.com/yongleex/SBCC}) is also available for interested researchers.

\end{abstract}

\maketitle


\section{Introduction}
\label{sec_intro}

\textit{Particle Image Velocimetry}~(PIV) is a popular non-intrusive instrument for flow field measurement in experimental fluid dynamics\cite{adrian1984scattering,raffel2018particle,lee2021diffeomorphic}. PIV generates quantitative vector field by analyzing the consecutive particle recordings. 
However, the particle images are easily deteriorated in a practical measurement due to non-uniform light illumination, light reflections, background noise sources, camera dark noise, etc~\cite{honkanen2005background,deen2010image,sciacchitano2014elimination}. Therefore, the accuracy and robustness of PIV results could be significantly decreased, recognized as \textit{uncertainty}~\cite{sciacchitano2019uncertainty}, \textit{peak-locking}~\cite{michaelis2016peak} and/or \textit{outliers}~\cite{wang2015proper,lee2017outlier}. Thus, in this work,
we focus on the challenging PIV estimation problem caused by the deteriorated particle recordings.

Over the past 40 years, the mainstream velocity estimation methods --- cross-correlation~\cite{willert1991digital,scarano2001iterative,wang2020globally,zhu2022approach,gao2021robust}, optical flow~(OF)~\cite{corpetti2006fluid,zhong2017optical,lu2021accurate} and deep neural-network~(DNN) regression~\cite{lee2017piv,cai2019dense,lagemann2021deep,cao2024three}--- are not particularly designed for robust PIV estimation. 
1). The vanilla standard cross-correlation~(SCC) computes the image similarity (dot product) as a function of the relative displacement. And SCC is not robust to image noise (such as, additive background noise, non-uniform illumination) because the noise is also correlated~\cite{eckstein2009digital}. Therefore, the generalized cross-correlation (GCC) methods improve signal-to-noise ratio of PIV cross-correlation via post-processing the correlation coefficients with different spectral filters, including phase correlation~(PC)~\cite{horner1984phase}, symmetric phase-only filter~(SPOF)~\cite{wernet2005symmetric}, robust phase correlation~(RPC)~\cite{eckstein2009digital} to name a few.  As a result, the GCC methods have achieved acceptable performance and have been extensively equipped for the majority of PIV software.
2). The optical flow~\cite{corpetti2006fluid} methods employ a preservation principle, namely that particle image brightness attribute does not change after a movement, to estimate the particle displacement. 
The risk of failure rises if the brightness preservation principle breaks, which often occurs with image noise in practical measurements. Thus, replacing the brightness attribute with other robust attributes~(image gradient, image phase) could be a straightforward modification~\cite{zhong2017optical}. Besides, a deliberated OF model with improved regularization term also contributes to accurate PIV estimation~\cite{bao2014fast,lu2021accurate}. Meanwhile, the complex OF models often come with heavy computation cost. 
3). As efficient inference methods, DNN-based regression methods\cite{lee2017piv,cai2019dense,zhang2020unsupervised,yu2021lightpivnet,lagemann2021deep,yu2023deep} have been attracting researchers' interest due to the powerful model capacity. However, the generalization of DNN for PIV depends on the noise type of training dataset~\cite{lagemann2022generalization}. 
Totally speaking, the CC, OF, DNN can be treated as bivariate operators that only take in two particle image frames, regardless of  the concrete noise signal of a measurement.  
Herein, we focus on the CC methods employed by most practitioners~\cite{kahler2016main,liberzon2016openpiv,xie2022spatiotemporal}.

\begin{figure}[ht!]
\centering
\subfloat[Standard PIV pipeline with background subtraction]{\includegraphics[width=3.2in]{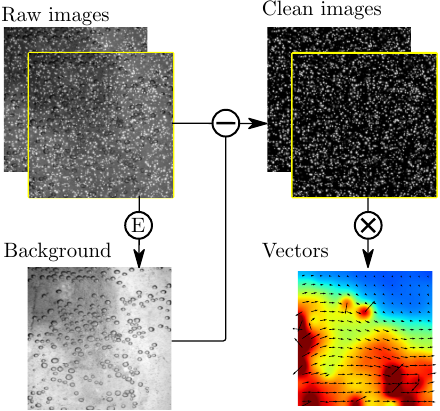}}%
\hfil
\subfloat[Our new PIV pipeline with SBCC]{\includegraphics[width=3.2in]{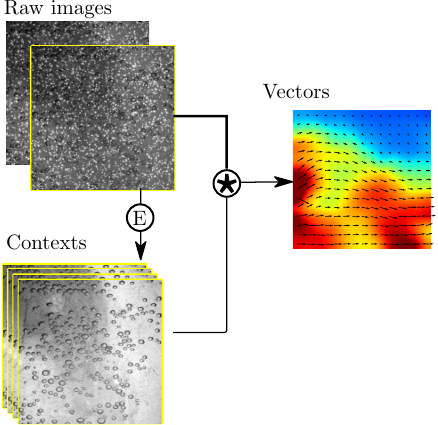}}%
\caption{Different illustrative pipelines for noisy PIV images. \raisebox{.5pt}{\textcircled{\raisebox{-.7pt} {E}}}: background/contexts extraction; $\ominus$: background subtraction; $\bigotimes$: cross-correlation. \raisebox{.5pt}{\textcircled{\raisebox{-.1pt} {$\star$}}}: surrogate-based cross-correlation.}
\label{fig_1}
\end{figure}

A straightforward alternative to achieve robust PIV analysis is to directly improve the image quality~\cite{dellenback2000contrast,shavit2007intensity,lee2020blind,fan2023deep,zhao2024flow}. 
Among different image pre-processing, the \textit{background subtraction} performs well given a good reference background--- concrete noise signal~\cite{mejia2013robust,mendez2017pod,kahler2016main,wang2020ratio}. 
The background image can either be recorded in the absence of seeding, or, if this is not possible, through temporal or spatial analysis from raw PIV recordings~\cite{raffel2018particle}.
The background image could be the minimum intensity image from double-frame PIV images, and works reliably for nonstational flow with severe background noise~\cite{honkanen2005background,deen2010image}.
The varying background image could be extracted via a temporal Butterworth filter from a large number of raw PIV recordings~\cite{sciacchitano2014elimination}. A customized background can be adaptive reconstructed through proper orthogonal decomposition (POD)~\cite{mendez2017pod}.
Without extra temporal information, the spatial low-pass  filter~(LPF) utilizes the blur image to approximate a background, including Gaussian filter, median filter~\cite{adrian2011particle}, anisotropic
diffusion~\cite{adatrao2019elimination}, etc.  Due to effectiveness, background subtraction has become an essential step of standard PIV pipeline~(Fig.~\ref{fig_1}(a)). However, using one background to model the complex noise signal is still challenging.

\textit{Correlation filters} have achieved competitive success in object tracking by learning a discriminative linear tracker from several image templates~\cite{bolme2010visual,henriques2012exploiting,henriques2014high}. 
It generates an optimal filter/tracker that maximizes the convolution/tracking performance from multiple templates, as detailed in Section~\ref{sec_related_works}. 
As a result of closed-form solution, correlation filter algorithm~(minimum output sum of squared error, MOSSE~\cite{bolme2010visual}) is not only easy to implement but significant faster. The extensive experiments exhibit that the learned filter outperforms the original image templates in both robustness and accuracy. 

Our insight is that taking more negative context images (the concrete noise signal) into account will obtain a more robust tracker,  and the MOSSE algorithm can be easily adapted to negative contexts with little effort. 
Different from object tracking video, the images of PIV pair are templates for each other~\cite{wereley2001second}.
As a result, a novel surrogate-based cross correlation (SBCC) framework is proposed by combining forward surrogate tracking with backward surrogate tracking. And SBCC---a multivariate operator--- enables us to reform PIV pipeline for accurate and robust measurement, rather than pursuing a perfect clean image~(Fig.~\ref{fig_1}(b)). The main contributions are:
\begin{enumerate}
  \item Inspired by the MOSSE filter, the SBCC is a new robust PIV analysis tool that employs several negative contexts, via robust surrogates and bi-directional consistency.
  \item Based on MOSSE objective and bi-directional consistency formula, a concise closed-form solution to the problem is obtained. To our surprise, a set of widely-used \textit{generalized cross-correlation} methods are special cases of the closed-form solution of SBCC framework. 
  \item The improvement of SBCC has been extensively verified on both synthetic and real PIV images. 


\end{enumerate} 

The rest of this paper is arranged as follows.
The related works are given in Section~\ref{sec_related_works}. Section~\ref{sec_sbcc}  describes our SBCC method from the problem formulation to optimization. And Section~\ref{sec_exp} demonstrates the performance on synthetic datasets and real PIV images in comparison with baseline methods. Finally, the Section~\ref{sec_conclusion} remarks the paper with several concluding comments.

\section{Related works}
\label{sec_related_works}

\begin{figure}[htb!]
\centering
\subfloat[SCC]{\includegraphics[width=3.0in]{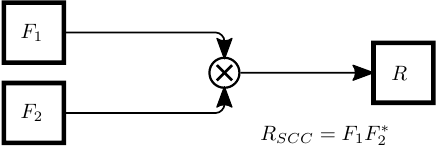}}%
\hfil
\subfloat[GCC]{\includegraphics[width=3.0in]{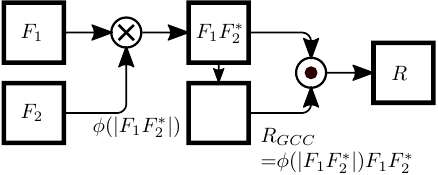}}%
\caption{Different cross-correlation methods. The $\bigotimes$ denotes the standard cross-correlation in Fourier frequency domain~(a). 
The generalized cross-correlation methods~(b), the $\bigodot$ represents an element-wise multiplication.}
\label{fig_2}
\end{figure}

\subsection{Cross-correlations}
The cross-correlation response $r(\mathbf{x})$ of image pair $f_1(\mathbf{x})$, $f_2(\mathbf{x})$ indicates the image displacement.
Due to the Convolution Theorem, the computation of cross-correlation $r(\mathbf{x})$ becomes a fast element-wise multiplication in the Fourier frequency domain. That says, $R(\mathbf{\omega}) = F_1(\mathbf{\omega}) F^*_2(\mathbf{\omega})
$, 
where $R(\omega), F_1(\omega), F_2(\omega)$ are the Fourier transform of $r(\mathbf{x})$, $f_1(\mathbf{x})$, $f_2(\mathbf{x})$ 
and the superscript $(^*)$ denotes complex conjugation.
Hereafter, we will simplify the notations ($F_1(\omega), F_2(\omega),...$) as ($F_1,F_2,...$) by omitting the frequency $\omega$. Due to its wide adoption in PIV estimation, this vanilla cross-correlation method will be referred to as the \textit{standard cross-correlation} (SCC) method, as shown in Fig.~\ref{fig_2}~(a).
\begin{equation}
R_{SCC} = F_1 F^*_2
\end{equation} 


To enhance the correlation signal, the \textit{generalized cross-correlation} (GCC) methods amend the SCC correlation $R_{SCC}$ with different spectral filters, i.e., 
\begin{equation}
R_{GCC} = \psi(F_1,F_2)F_1F_2^*
\label{eq_gcc}
\end{equation}
where $\psi(\cdot)$ denotes the modification operation (PHAT filter~\cite{horner1984phase}, SPOF filter~\cite{wernet2005symmetric}, RPC filter~\cite{eckstein2009digital}, etc.). Compared with the $R_{PRE}$ (image pre-processing),  the $R_{GCC}$ can be viewed as a post-processing of $F_1F_2^*$. 
Several GCC instances are listed in Table.~\ref{tab_1}. The filters of these GCC methods share a special type, $\psi(F_1,F_2)=\phi(|F_1F_2^*|)$, as demonstrated in Fig.~\ref{fig_2}~(b). 
%


\begin{table}
\centering
\caption{Different cross-correlation methods. }
\label{tab_1}       
\begin{tabular}{lll}
\hline\noalign{\smallskip}
\textbf{Methods} &  \textbf{Equations} &  \textbf{Comments} \\
\noalign{\smallskip}\hline\noalign{\smallskip}
SCC\cite{raffel2018particle} & $R_{SCC} = F_1F_2^*$ & -  \\
\hline\hline
Pre-processing & $R_{PRE} = (HF_1)(HF_2)^*$ & Filter $H$ \\
\hline
Background subs & $R_{BGS} = (F_1-B)(F_2-B)^*$ & Background $B$ \\
\hline
\hline
PC\cite{horner1984phase}       & $R_{PC} = \frac{F_1F_2^*}{|F_1F_2^*|}$ &  GCC method \\ 
SPOF\cite{wernet2005symmetric}     & $R_{SPOF} = \frac{F_1F_2^*}{\sqrt{|F_1F_2^*|}}$ &  GCC method \\
RPC\cite{eckstein2008phase,eckstein2009digital}      
& $R_{RPC} = \frac{GF_1F_2^*}{|F_1F_2^*|}$ &  GCC method \\
$\rho$-CSPC \cite{shen2009modified} & $R_{CSPC} = \frac{F_1F_2^*}{|F_1F_2^*|^{\rho}+\epsilon}$ &  GCC method\\
\hline
\hline
\textbf{SBCC} (ours) & $R_{SBCC} =  \frac{2GF_{1} F_{2}^*}{F_{1}F_{1}^*+F_{2}F_{2}^* + 2\mu \Sigma P_i P_i^*}$ & Contexts $P_i$ \\
\noalign{\smallskip}\hline
\end{tabular}
\end{table}

\subsection{Correlation filter}


The correlation filter can be derived either from an objective function specifically formularized in the Fourier domain~\cite{bolme2010visual} or from ridge regression and circulant matrices~\cite{henriques2012exploiting}. 
Slightly different from~\cite{bolme2010visual,henriques2012exploiting}, we provide our understanding of correlation filter as following. 
Given a set of aligned template images $T_i,i\in \{1,2,...,n\}$, the MOSSE method~\cite{bolme2010visual,henriques2012exploiting} finds a surrogate filter/tracker $S$ that produces the best tracking performance,
i.e., the cross-correlation response $r_i(\textbf{x}) = \mathcal{F}^{-1}(T_i S^*)$ is encouraged to be an isotropic Gaussian-shaped response $g(\mathbf{x})$, where $\mathcal{F}^{-1}(\cdot)$ denotes inverse fourier transform. 
In addition, a regularization term $|S|^2$ is employed to avoid the over fitting and gains the stability, similar to Wiener filtering or ridge regression~\cite{henriques2012exploiting}. Hence, 
%
the minimum output sum of squared error (MOSSE) objective arrives, 
\begin{equation}
\begin{split}
{J}_{MOSSE}(S) = \mathop{\Sigma}_{i=1}^n |G-T_i S^*|^2 + \mu |S|^2
\label{Eq_mosse}
\end{split}
\end{equation}
where $\mu$ controls the amount of regularization, and it is recommended to be $0.1$~\cite{bolme2010visual}. The $n$ is the number of positive templates, and $G$ denotes the Fourier transform of a Gaussian function $g(\mathbf{x})$. A closed-form solution of MOSSE, 
\begin{equation}
\hat{S}^* = \frac{\Sigma_i G T_i^*}{\Sigma_i T_i T_i^*+\mu}
\end{equation}
The $\mu$ plays an important role to make the denominator not equal to zero. Interestingly, the numerator is the correlation between the input and the desired output and the denominator is the energy spectrum of the
input.

Our observation is that the regularization term of Eq.(\ref{Eq_mosse}), $|S|^2=|0-1\cdot S^*|^2$, could be regarded as a special term for a negative template (delta function, $1\Leftrightarrow \delta(\mathbf{x})$). That is to say, the MOSSE filter also expects the cross-correlation between $S$ and a negative Dirac delta function to be zero. Now, it is clear that the parameter $\mu$ controls the relative importance for this negative template response.
 
\section{Surrogate-based cross-correlation}
\label{sec_sbcc}

\subsection{Problem formulation} 
\label{problem_formulation}

\begin{figure}[ht]
\centering
\includegraphics[width=3.0in]{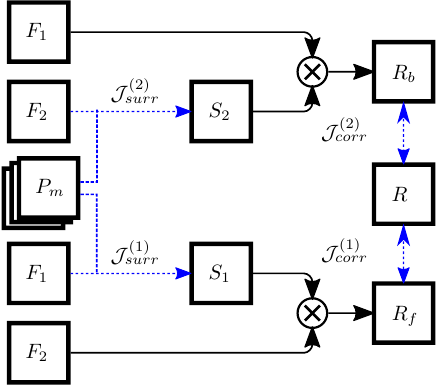}
\caption{The bi-directional surrogate model, SBCC. The surrogate images $S_1, S_2$ and correlation response $R$ are jointly optimized with surrogate objective ${J}_{surr}$ and correlation consistency objective ${J}_{corr}$ (blue dash arrows). The $R_f$ and $R_b$ represent the correlation with forward surrogate ($S_1$) and backward surrogate ($S_2$) respectively.}
\label{fig_3}
\end{figure}

As mentioned in Section~\ref{sec_intro}, the standard PIV pipeline might fail due to one limited background image.
We thus introduce the surrogate-based cross-correlation which utilizes multiple context background images to enhance particles correlation. Specifically, a surrogate filter/image ($S_1$ or $S_2$) is assumed to have a better cross-correlation response under a well-designed surrogate objective $J_{surr}$, which considers the tracking performance as well as robustness.
Meanwhile, similar to ensemble correlation, the forward correlation response $R_f$ (with surrogate $S_1$) and backward response $R_b$ (with surrogate $S_2$) are combined via correlation consistency objective ${J}_{corr}$. Thus, the robust PIV estimation problem is formulated with two objectives, as illustrated in Fig.~\ref{fig_3},

\begin{equation}
\hat{R}, \hat{S}_1, \hat{S}_2 = \mathop{\arg\min}_{R,S_1,S_2} {J}_{SBCC}(R, S_1, S_2; F_1,F_2)
\end{equation} 
with 

\begin{equation}
\begin{split}
{J}_{SBCC}& (R, S_1, S_2; F_1,F_2)\\
 &= \underbrace{{J}_{surr}(S_1;F_1)+{J}_{surr}(S_2;F_2)}_\textrm{surrogate objective}\\
&\quad +\underbrace{{J}_{corr}(R_f,R)+{J}_{corr}(R_b,R)}_\textrm{correlation consistency  objective}
\end{split} 
\label{eq_objective_SBCC}
\end{equation}
where $R_{f} =S_1F_2^*, R_{b}=F_1S_2^*$ are the forward correlation response and backward correlation response. 
Note that the surrogates $S_1, S_2$ are no longer the processed results of image pre-processing due to the coupled SBCC structure. 

\begin{figure}[!htbp]
\centering
\includegraphics[width=3.0in]{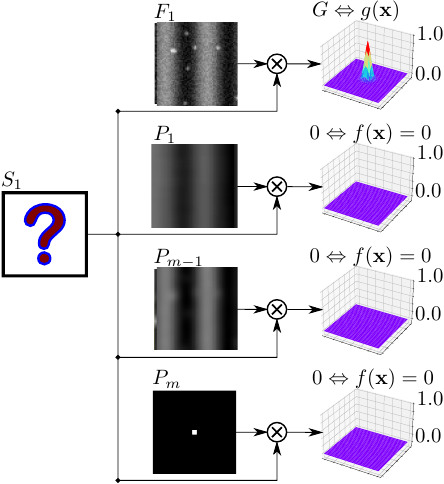}
\caption{The motivation of surrogate objective is to find a surrogate $S_1$ that does not response to the negative context images $P_1, P_2, ..., P_m$.
}
\label{fig_4}
\end{figure}

\textit{Surrogate objective.} 
To gain robustness of surrogate filter, a well-designed surrogate objective is thus constructed, which makes use of the positive template and other negative context images. The negative samples are proved to be useful for representation learning~\cite{Chen2021MuMMI}. Similar to MOSSE~\cite{bolme2010visual}, our surrogate objective ${J}_{surr}$ (Fig.~\ref{fig_4}) is given before a detail explanation,

%
\begin{equation}
\label{eq_7}
\begin{split}
{J}_{surr}(S; F) &= \underbrace{|G-FS^*|^2}_\textrm{MOSSE term}+\underbrace{\mu \frac{1}{m}\mathop{\Sigma}_{i=1}^m|0-P_i S^*|^2}_\textrm{negative context term}\\
&= |G-FS^*|^2 + \mu \frac{1}{m}\mathop{\Sigma}_{i=1}^m|P_i S^*|^2\\
\end{split}
\end{equation}
where $P_i, i\in\{1,2,...,m\}$ are the negative context images. The $G$ denotes the Fourier transform of Gaussian functions $g(\mathbf{x})$. The $\mu$ is also the coefficient for the negative context term.
The MOSSE term is preserved to encourage a Gaussian-shaped correlation map.
%
Reducing the filter's response to the background/noise could decrease the number of outliers, because most outliers occur when the images have similar image background or other noisy pattern. Recall that the regularization term $|S|^2$ of MOSSE, it encourages the surrogate filter $S$ produce zero response to the special $\delta(\mathbf{x})$. However, the negative context term of Eq.~(\ref{eq_7}) encourages the filter $S$ produce zero response to all context images. 
We choose the backgrounds from temporal minimum value (MIN bg) and spatial low-pass results (LPF bg) as the context images in this work. Obviously, other options are also supported. Compared to the $\delta(\mathbf{x})$, the context images are  more likely to have a similar noisy pattern with PIV test images. 

\textit{Correlation consistency objective}. 
Different from object tracking, the paired images of PIV can be treated as templates to each other. Our SBCC framework (Fig.~\ref{fig_3}) models it as forward and backward correlation. To obtain a consistency result, a square error encourages a minimum distance between $R_x\in \{R_b, R_f\}$  to the final correlation map $R$.
\begin{equation}
\label{eq_10}
\begin{split}
{J}_{corr}(R_x,R) &= |R-R_x|^2
\end{split}
\end{equation}
Recall $R_{f} =S_1F_2^*, R_{b}=F_1S_2^*$ are the forward correlation response and backward correlation response, and $R$ is the final cross-correlation response of SBCC.

Take the \textit{surrogate objective} (Eq.~(\ref{eq_7})) and \textit{correlation consistency objective} (Eq.~(\ref{eq_10})) back into the problem (Eq.~(\ref{eq_objective_SBCC})). The specific objective function of SBCC is arrived,
\begin{equation}
\label{eq_9}
\begin{split}
{J}_{SBCC} &(R, S_1, S_2; F_1,F_2)\\
 &= |G-F_1S_1^*|^2+\mu \frac{1}{m}\mathop{\Sigma}_{i=1}^m|P_i S_1^*|^2\\
 &\quad + |G-F_2S_2^*|^2+\mu \frac{1}{m}\mathop{\Sigma}_{i=1}^m|P_i S_2^*|^2\\
&\quad + |R-S_1F_2^*|^2+|R-F_1S_2^*|^2\\
\end{split} 
\end{equation}
This objective is the sum of several squared errors with three unknown complex variables $S_1, S_2, R$. The parameter $\mu$ controls the relative importance of the negative context images.

\subsection{Optimization of SBCC objective}
The optimization of ${J}_{SBCC}$ (Eq.~\ref{eq_9}) is almost identical to the optimization problems in \cite{bolme2010visual,henriques2012exploiting}. The difference is that SBCC objective is a quadratic convex function with three complex variables.  
The closed-form solution thus can be found by setting the partials to zeroes,


\begin{equation}
\label{eq_sbcc_solution}
\begin{split}
R_{SBCC}:=\hat{R}&=\frac{2GF_{1} F_{2}^*}{F_{1}F_{1}^*+F_{2}F_{2}^*+2\mu Q}\\
\hat{S}_{1} 
    &= \frac{ F_{2}\hat{R}+GF_{1} }{F_{1}F_{1}^*+F_{2}F_{2}^* +\mu  Q}\\
\hat{S}_{2}^* 
    &= \frac{ F_{1}^* \hat{R} +G F_{2}^*}{F_{1}F_{1}^*+F_{2}F_{2}^* +\mu  Q} \\
\end{split}
\end{equation}
where $Q=\frac{1}{m}\Sigma_{i=1}^m P_iP_i^*$ is the average Fourier power spectrum of negative context images. The cross-correlation response, $R_{SBCC}$, incorporates this $Q$ component to obtain a robustness correlation by considering the noisy background in these negative context images. 
The terms in $R_{SBCC} ($Eq.~(\ref{eq_sbcc_solution})) have clear interpretation. The numerator is the cross-correlation between $F_1$ and $F_2$ with a Gaussian filter ($G$), and the denominator is the power spectrum sum of $F_1$, $F_2$, and negative context images $P_i$. Note that the $\hat{S}_1$ depends on $F_1$ as well as $F_2$.

\subsection{Short discussion}
Due to the clear meaning of the objective function ${J}_{SBCC}$, the solution $R_{SBCC}$ has good interpretation as well.
Observing the SBCC solution, we found that a set of hand-crafted GCC methods are special cases of SBCC. That is to say, our SBCC framework provides a new perspective to understand existing GCC methods (Table.~\ref{tab_1}).

Firstly, we consider a simple but useful equation, 
\begin{equation}
F_1F_1^* + F_2F_2^* = (|F_1|-|F_2|)^2+2|F_1F_2^*|\ge 2|F_1F_2^*|
\end{equation}
Thus, the mathematical representation of PC, RPC, 1-CSPC become special SBCC cases with an impractical noise-free assumption ($|F_1|=|F_2|$). That says, 
\begin{equation}
\begin{split}
R_{PC} \ \  &= R_{SBCC}|_{G=1, |F_1|=|F_2|, \mu=0}\\
R_{1-CSPC}  &= R_{SBCC}|_{G=1, |F_1|=|F_2|, \mu=\epsilon, P_i=1}\\
R_{RPC} &= R_{SBCC}|_{G\Leftrightarrow g(x), |F_1|=|F_2|, \mu=0}\\
\end{split}
\end{equation}
which means that, PC and 1-CSPC encourage a Dirac delta response $(G=1)$, while RPC method expects a Gaussian-shaped response.  Only 1-CSPC method has considered a special context image~$\delta(\mathbf{x})$ implicitly. However, all methods employ a noise-free assumption~($|F_1|=|F_2|$).

The SPOF is beyond a brief understanding. However, the SPOF can be treated as an ensemble correlation~\cite{delnoij1999ensemble} due to the following relationship, $R_{SPOF}R_{SPOF}=R_{SCC}R_{PC}$.
It might imply that multiple SBCC frameworks can provide more complex surrogate for SPOF and general $\rho-$CSPC methods. Note that none of the existing GCC methods explicitly take any negative context images into consideration.

\section{Experiments}
\label{sec_exp}
 
In this section, the performance of SBCC is extensively evaluated through visualizing the correlation map, analysing parameter sensitivity, conducting measurement on synthetic images and experimental PIV images. The detailed implementation and additional results are provided at the project repository, \url{https://github.com/yongleex/SBCC}.


\textit{Baseline methods.} Several widely-accepted approaches are adopted to conduct a fair evaluation. And they are, standard cross-correlation~(SCC)~\cite{raffel2018particle}, symmetric phase-only filter~(SPOF)~\cite{wernet2005symmetric}, robust phase correlation~(RPC)~\cite{eckstein2009digital}. Regarding the background subtraction methods, we choose the minimum intensity image from double-frame PIV image~\cite{honkanen2005background} and the spatial low-pass filter (LPF)~\cite{adrian2011particle}, resulting in SCC-MIN and SCC-LPF. To exclude the influence of other factors, single-pass cross-correlation without any post-processing is utilized for all testing methods.

\textit{Evaluation criteria.} In addition to subjective visual judgement, three objective metrics are also employed to quantify the performance: 1) the root mean-square-error (RMSE)~\cite{raffel2018particle,lee2017piv,lee2021diffeomorphic}, 2) the average endpoint error (AEE)~\cite{lagemann2021deep}, 3) the execution time for different image size. 
\begin{equation}
\begin{split}
  RMSE&=\sqrt{\frac{1}{N}\mathop{\Sigma}_{i=1}^N \|\mathbf{v}_{e,i}-\mathbf{v}_{t,i}\|^2}\\
  AEE&=\frac{1}{N}\mathop{\Sigma}_{i=1}^N \|\mathbf{v}_{e,i}-\mathbf{v}_{t,i}\|
\end{split}
\end{equation}
where $\mathbf{v}_{e,i}=(v_x,v_y)$ is the $i^{th}$ estimated vector out of $N$ points, while the $\mathbf{v}_{t,i}$ denotes the $i^{th}$ ground truth.

\subsection{On correlation coefficients}


\begin{figure}[!htbp]
\centering
\subfloat[A test image pair $(u=-5.0 pixel)$ and estimated backgrounds]{
\includegraphics[width=\columnwidth]{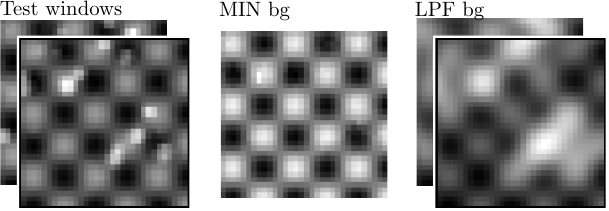}}%
\hfil
\subfloat[The correlation maps]{
\includegraphics[width=\columnwidth]{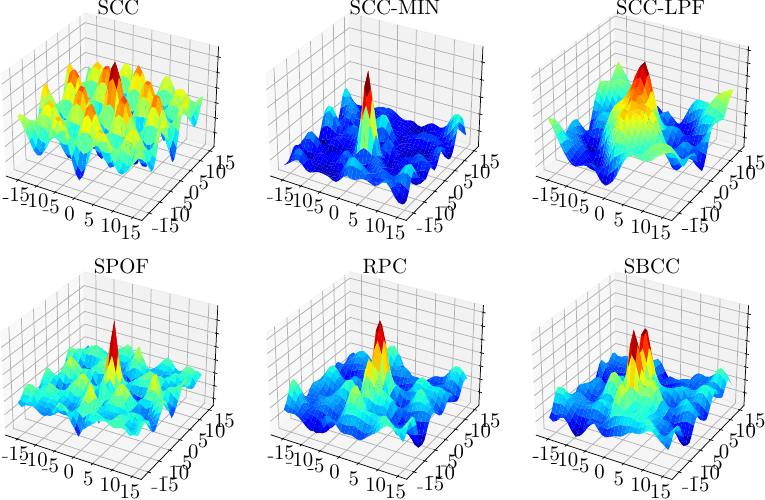}}%
\hfil
\subfloat[The correlation coefficient at $\Delta y=0$]{
\includegraphics[width=\columnwidth]{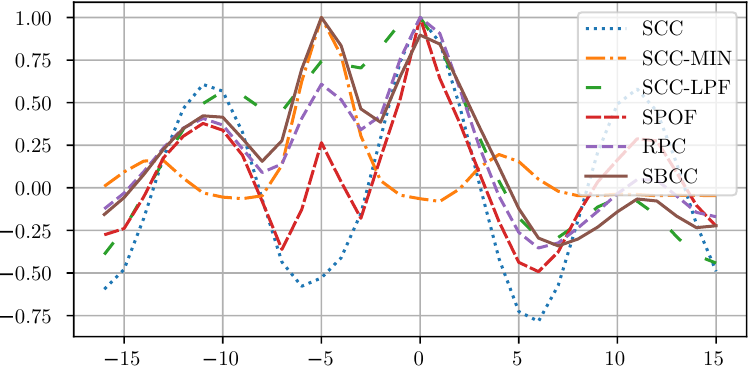}}
\caption{A correlation responses on synthetic particle images with background. Best viewed in color.}
\label{fig_5}
\end{figure}

Fig.~\ref{fig_5}~(a) gives a test pair of synthetic PIV interrogation window (particle displacement is $-5.0pixel$ in horizontal), with a unrealistic strong additive background.  Obviously, the minimum intensity image~(MIN bg) recovers the still background, while the LPF~(LPF bg) can not tell the background and particle image apart due to frequency aliasing. 

Fig.~\ref{fig_5}~(b) and (c) provide the cross-correlation coefficients for this challenging synthetic case. The SCC, SPOF, RPC fail to obtain the correct response peak due to a lack of noise signal. 
Although the SCC-LPF method does not show the correct peak, the rough background also helps to increase the image similarity at the correct displacement. 
Without surprising, the SCC-MIN method is demonstrated with perfect correlation peak with good background estimation in this case.
Compared to other methods, the SBCC has a correlation map with two distinct peaks, and correlation peak (maximum similarity) is correctly located at $(-5.0,0.0)$. 
Closing observing the curve around $(-5.0,0.0)$, the SBCC has a similar landscape with SCC-MIN. Thus, we can conclude that SBCC provides another feasible mechanism to perform robust cross-correlation with the help of background signals.

\subsection{On parameter sensitivity}

The only parameter $\mu$ needs to be determined for our SBCC method working at the best condition. Similar to the background subtraction, it's very difficult to obtain ideal context images. Hence, increasing $\mu$ will not always benefit the robustness or accuracy. We thus argue that there is an optimal parameter $\mu$ for the practical PIV measurement.

\begin{figure}[htbp]
\centering
\includegraphics[width=\columnwidth]{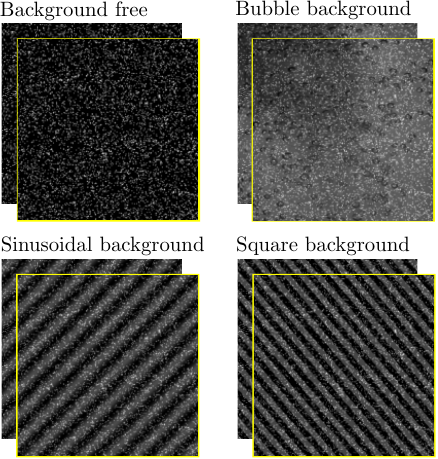}
\caption{Synthetic particle images with/without different backgrounds~(uniform flow).}
\label{fig_6}
\end{figure}

To study the effect of different $\mu$ values, we also employ the synthetic particle images from particle image generator~(PIG). Here, we use a subset of a ready dataset (1000 uniform flow)~\cite{cai2019particle,cai2019dense}.  
Different from the noise-free images, a random synthetic image is added to the clean synthetic image pair, to simulate the still background. The background images are from a bubble dataset (1219 images of class 1)~\cite{bai2021classification,buble2021}, synthetic sinusoidal and square signal, as demonstrated in Fig.~\ref{fig_6}.

\begin{figure}[htbp]
\centering
\subfloat[RMSE for a background-free image pair]{
\includegraphics[width=\columnwidth]{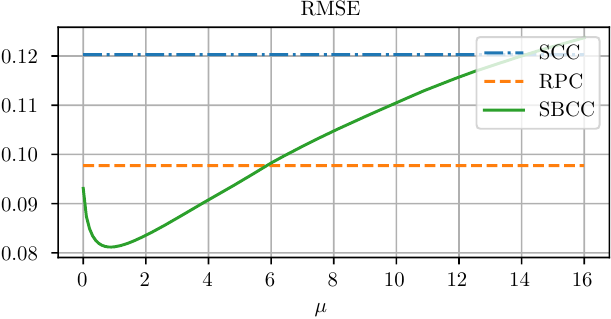}}%
\hfil
\subfloat[RMSE for an image pair with bubble background]{
\includegraphics[width=\columnwidth]{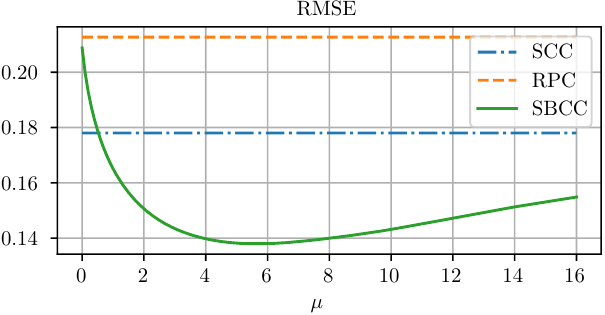}}
\caption{Effect of the parameter $\mu$ of SBCC.~($u=0.85 pixel, v=1.23 pixel$)}
\label{fig_7}
\end{figure}

Fig.~\ref{fig_7} illustrates the experimental results with both background- free and bubble background image pair~(top row of Fig.~\ref{fig_6}) respectively, and the parameter $\mu$ varies from 0 to 16 with interval $0.1$. 
In comparison with noise-free case, the RMSE of all methods are increased when additive bubble noise is added. It reflects the PIV challenge caused by the unwanted background. 
Interestingly, the sensitivity curve of SBCC have an optimal value that corresponds with the minimum RMSE value.
The optimal value $\mu$  of complex background is larger than that of background-free situation. 
Meanwhile, an improvement of SBCC happens for a large range of $\mu$, i.e., range $[0,6]$ for background-free and $[1,16]$ for bubble background. Thus, setting a proper $\mu$ could be an interesting problem for future work. Anyway, we set $\mu$ to $3.0$ arbitrarily by taking all cases into consideration based on the results of this experiment. Note that, the fixed $\mu(3.0)$ is not changed for different measurement cases, and the extensive results illustrate that this value could have a universal robust performance.

\subsection{On synthetic PIV images}

\begin{table}[htbp]
\centering
\caption{Synthetic experiment on one image pair. Performance measured by RMSE. The best \textbf{in Bold}.}
\label{tab_2} 
\begin{tabular}{|c|c|c|c|c|}
\hline
\textbf{Background}         & Free & Bubble & Sinusoidal & Square  \\
\hline  \hline
\textbf{SCC}          &  0.3753  &  0.5054  &  0.7930  &  0.4644  \\ \hline
\textbf{SCC-MIN}      &  0.4422  &  0.4422  &  0.4422  &  0.4422  \\ \hline
\textbf{SCC-LPF}      &  0.3919  &  0.3487  &  0.3934  &  0.3244  \\ \hline
\textbf{SPOF}         &  0.4060  &  0.4530  &  0.4725  &  0.4382  \\ \hline
\textbf{RPC}  &  \textbf{0.1883} &  0.6597  &  0.3724  &  0.2274  \\ \hline
\textbf{SBCC}         &  0.2195  &  \textbf{0.2765} &  \textbf{0.1914} & \textbf{0.1985} \\ \hline
\end{tabular}
\end{table}

\begin{table}[!htbp]
\centering
\caption{Synthetic experiment on one image pair. Performance measured by AEE. The best \textbf{in Bold}.}
\label{tab_3} 
\begin{tabular}{|c|c|c|c|c|}
\hline
\textbf{Background}         & Free & Bubble & Sinusoidal & Square  \\
\hline \hline
\textbf{SCC}          &  0.3422  &  0.4557  &  0.7813  &  0.4311  \\ \hline
\textbf{SCC-MIN}      &  0.3827  &  0.3827  &  0.3827  &  0.3827  \\ \hline
\textbf{SCC-LPF}      &  0.3438  &  0.3213  &  0.3721  &  0.3054  \\ \hline
\textbf{SPOF}         &  0.3768  &  0.4181  &  0.4460  &  0.4099  \\ \hline
\textbf{RPC}  &  \textbf{0.1619} &  0.4193  &  0.3440  &  0.1983  \\ \hline
\textbf{SBCC}         &  0.2050  &  \textbf{0.2457} &  \textbf{0.1741} & \textbf{0.1810} \\ \hline
\end{tabular}
\end{table}

To quantitatively compare the performance, a synthetic image pair is sampled~\cite{cai2019dense} with different backgrounds (Fig.~\ref{fig_6}). Table.~\ref{tab_2} and \ref{tab_3} give the RMSE and AEE values for the processed results.  For the background-free scenario, vanilla SCC yields satisfactory result, while the variations (SCC-MIN, SCC-LPF and SPOF) do not show a consistent improvement in accuracy.
For the cases with background, all variants have a better performance than SCC. It implies that background subtraction (SCC-MIN, SCC-LPF) and spectral filters (SPOF, RPC) are all effective to address background problem. 
In this experiment, the RPC has the smallest measurement error on background-free images, 
while SBCC achieves significant improvement for the cases with background.

\begin{figure}[!ht]
\centering
\includegraphics[width=3.2in]{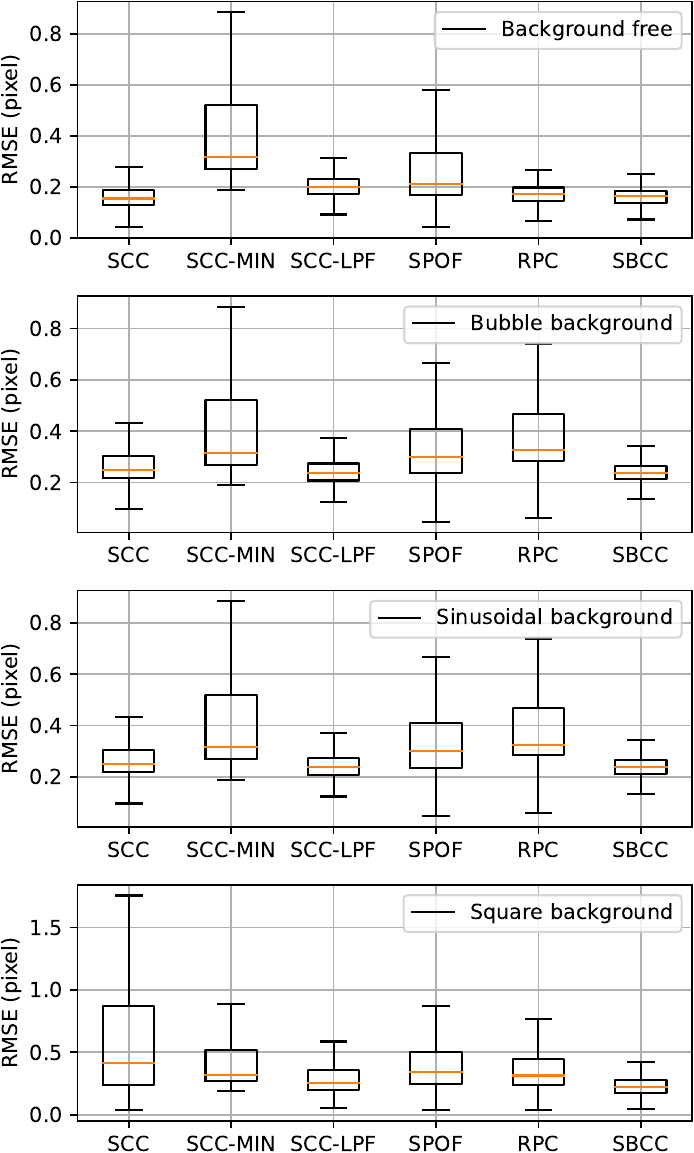}
\caption{Boxplot of the RMSE for the synthetic 1000 uniform cases with 4 types of backgrounds and 6 PIV CC estimators.}
\label{fig_8}
\end{figure}

\begin{figure}[!ht]
\centering
\includegraphics[width=3.2in]{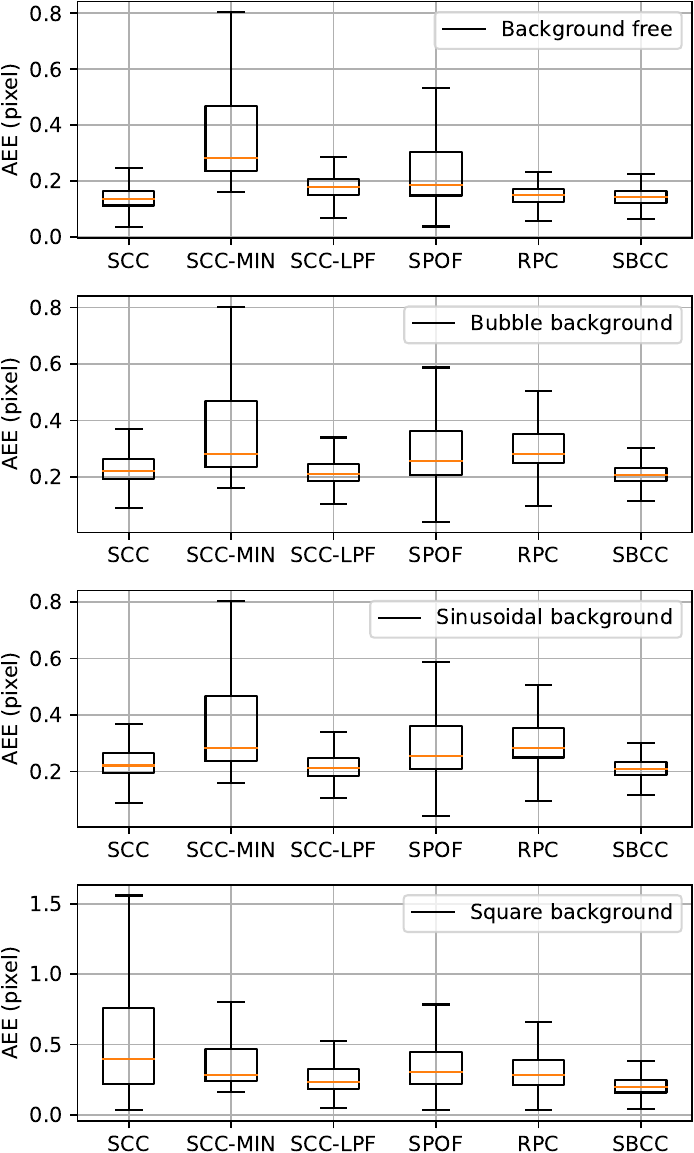}
\caption{Boxplot of the AEE for the synthetic 1000 uniform cases with 4 types of backgrounds and 6 PIV CC estimators.}
\label{fig_9}
\end{figure}

In addition to case study, the Monte Carlo simulation is widely adopted in the assessment of PIV measurement uncertainty~\cite{raffel2018particle,lee2017piv}. 
Recall that, the 1000 synthetic particle image pairs ($256\times 256$ pixel$^2$) are synthesized with uniform flows ground truth~\cite{cai2019particle}, and the backgrounds are sampled from a bubble dataset~\cite{buble2021} and two artificial signals (random sinusoidal wave and square wave). The boxplots in Fig.~\ref{fig_8} and \ref{fig_9} present the statistical results of 1000 cases measured by RMSE and AEE. The one-pass SCC and RPC method have acceptable measurement error (RMSE $\sim 0.2$) for background-free cases due to varying seeding density, particle diameters, illuminations in the image generator. However, they are not robust enough to backgrounds. Both SCC-MIN and SPOF have poor performance on all cases. The reasons might be the low-quality background of SCC-MIN, while peak-locking causes a significant error of SPOF. On the contrary, both SCC-LPF and SBCC performs well for all test cases. We speculate that the synthetic backgrounds might be well estimated by a LPF, resulting in the good performance of SCC-LPF. Note that, our SBCC is more accurate than SCC-LPF statistically.  

\subsection{On real recorded PIV images}

\begin{figure}[]
\centering
\subfloat[A lab-made PIV setup with an industrial camera (HIKROBOT, MV-CA016-10UM) and multiple semiconductor lasers~(532$nm$)]{\includegraphics[width=0.6\columnwidth]{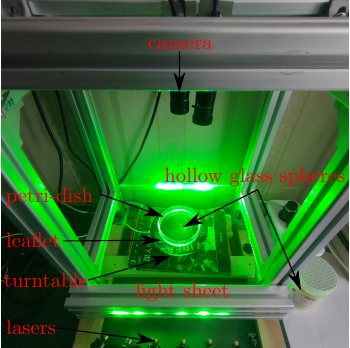}}
\hfil
\subfloat[Left: interactive flow around L-shaped plate~\cite{liberzon2016openpiv}; Middle: hypersonic flow over a step model~\cite{lu2023research}; Right: rotational flow with background from (a).]{\includegraphics[width=\columnwidth]{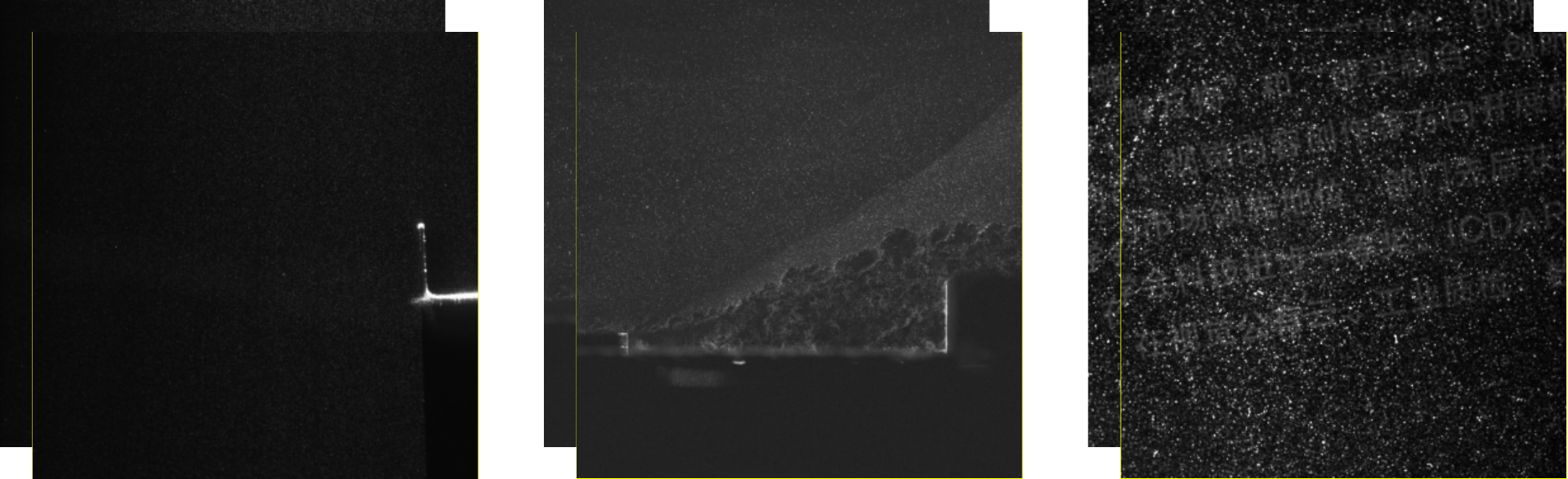}}
\caption{A lab-made setup~(a) and 3 practical test cases~(b).}
\label{fig_10}
\end{figure}

\begin{figure*}[]
\centering
\includegraphics[width=6.0in]{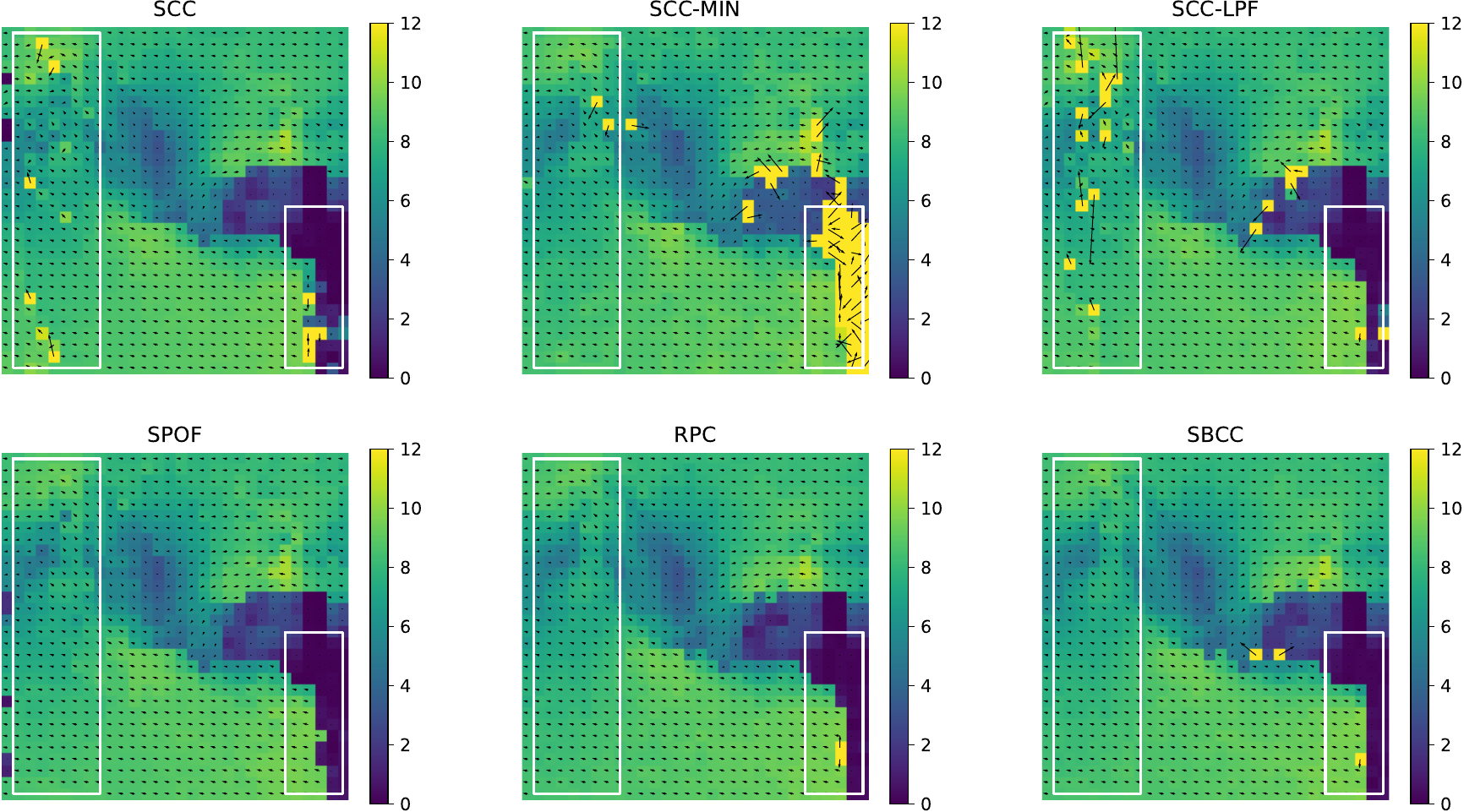}
\caption{Vectors from the interactive flow. The color background corresponds to velocity magnitude. Best viewed in color.}
\label{fig_11}
\end{figure*}

\begin{figure*}[!t]
\centering
\includegraphics[width=6.0in]{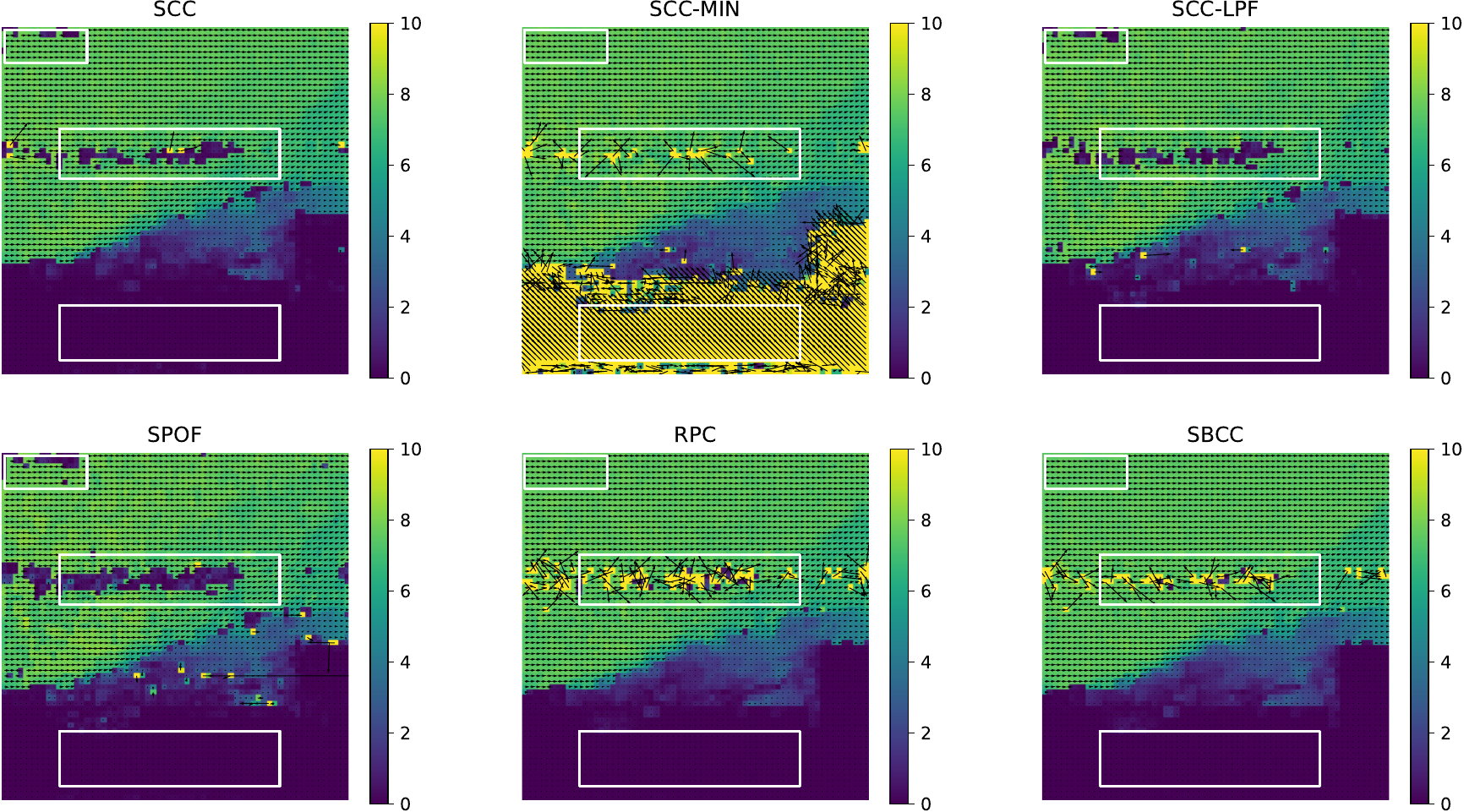}
\caption{Vectors from the hypersonic flow. The color background corresponds to velocity magnitude. Best viewed in color.}
\label{fig_12}
\end{figure*}

\begin{figure*}[!ht]
\centering
\includegraphics[width=6.0in]{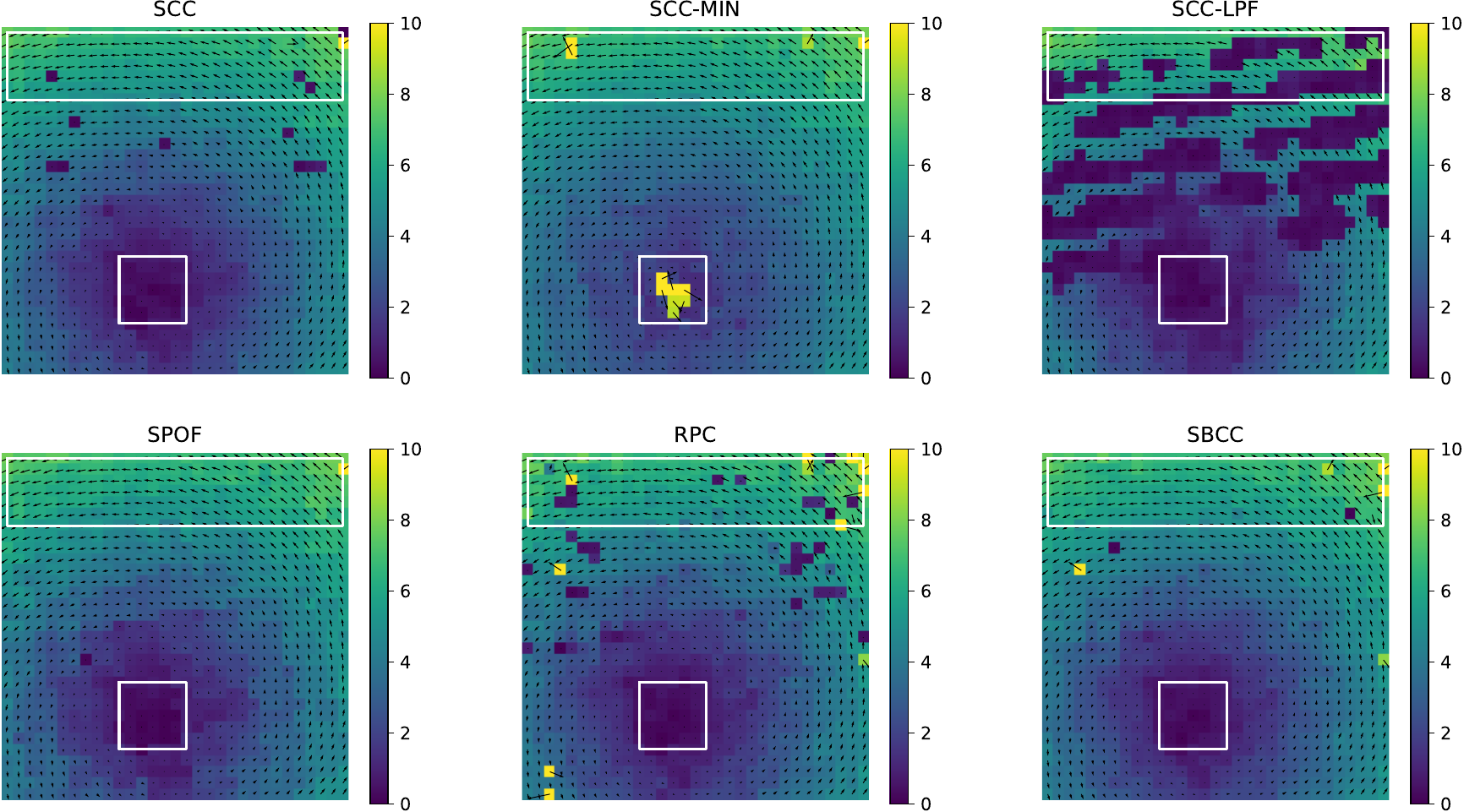}
\caption{Vectors from the rotational flow. The color background corresponds to velocity magnitude. Best viewed in color.}
\label{fig_13}
\end{figure*}

In addition to synthetic images, we also tested SBCC on three challenging recorded PIV image pairs (Fig.~\ref{fig_10} (b)). The first case records an interactive flow around L-shaped plate in OpenPIV~\cite{liberzon2016openpiv}. The second PIV case records a hypersonic flow ($5Ma$) over a step model~\cite{lu2023research}. The third image pair is from a lab-made PIV setup (Fig.~\ref{fig_10} (a)), which represents a liquid column rotating with a still text leaflet background (similar to PIV Challenge 2014, Case F). These flows have particular structures with large displacement ($\approx 10$ pixel), and the images have non-uniform illumination, out-of-plane effects and noise background. Thus, they are considered as the challenging test examples.


Fig.~\ref{fig_11}, \ref{fig_12} and \ref{fig_13} provide the raw vector field results, computed with different one-pass cross-correlation methods without any post-processing operation. The pseudo-color backgrounds represent the vector magnitudes. Overall, different methods output similar flow patterns, verifying that these widely-used methods do work in practice. Unfortunately, the accuracy for each method can not be exactly assessed due to unknown ground truth. 
Thus, we visually check the outliers for the problematic areas in white boxes to evaluate the robustness of each method.
The left area of interactive flow image is under weak illumination, and thus full of random noise. The results indicate the spectral filters (SPOF, RPC and SBCC) can effectively cope with this problem, as reported in related works~\cite{eckstein2009digital}. 
The middle box of hypersonic flow, with strong out-of-plane movement, is full of uncorrelated particles. We argue that the MIN background could catch some content of the noise signal, resulting in the less outliers of SCC-MIN and SBCC.      
The complex background of rotational flow is obviously difficult to reconstruct with MIN or LPF approaches, which explains the bad performance of SCC-MIN and SCC-LPF. Besides, the spectral filter of RPC also does not work well for this case. However, combining multiple background contexts and spectral filter, our SBCC has an impressive performance with a few outliers for this strong background of advertising words. 
Based on the impressive results of comprehensive (synthetic and recorded) experiments, the overall effectiveness of our SBCC has been confirmed. 
 
\subsection{On computing cost}

\begin{table*}[!htbp]
\centering
\caption{The running time (seconds) (averaged over 10 runs, with standard deviations).}
\label{tab_4} 
\begin{tabular}{|c|c|c|c|c|}
\hline
\textbf{Image size}   & $128\times128$      & $256\times256$      & $512\times512$      &  $1014\times1024$  \\
\hline \hline
\textbf{SCC}          &  $0.0022\pm0.0002$  &  $0.0055\pm0.0009$  &  $0.0233\pm0.0019$  &  $0.0792\pm0.0031$  \\ \hline
\textbf{SCC-MIN}      &  $0.0020\pm0.0005$  &  $0.0055\pm0.0007$  &  $0.0220\pm0.0009$  &  $0.0785\pm0.0026$  \\ \hline
\textbf{SCC-LPF}      &  $0.0070\pm0.0127$  &  $0.0081\pm0.0008$  &  $0.0277\pm0.0006$  &  $0.0838\pm0.0030$  \\ \hline
\textbf{SPOF}         &  $0.0019\pm0.0001$  &  $0.0071\pm0.0004$  &  $0.0232\pm0.0005$  &  $0.0834\pm0.0032$  \\ \hline
\textbf{RPC}          &  $0.0022\pm0.0002$  &  $0.0067\pm0.0002$  &  $0.0232\pm0.0007$  &  $0.0822\pm0.0024$  \\ \hline
\textbf{SBCC}         &  $0.0070\pm0.0034$  &  $0.0187\pm0.0004$  &  $0.0576\pm0.0034$  &  $0.1995\pm0.0315$  \\ \hline
\end{tabular}
\end{table*}

To demonstrate the efficiency of proposed SBCC, the baseline methods and SBCC were finally tested using Python 3.8 on a 2.70GHz i5-11400H laptop computer (HP OMEN 16) with RAM 16.00GB. Tested over 10 runs, the execution results (Table.~\ref{tab_4}) with varied image size demonstrated that the SBCC has an acceptable computational cost. 
Recall that an extra FFT operation of contexts is needed, and it is worthwhile to cost $2\sim3$ times computation for the accuracy improvement. For a $1024 \times 1024$ case, the execution time of our SBCC is less than $0.2$ second.

\section{Conclusion}
\label{sec_conclusion}
Inspired by correlation filter, a novel SBCC framework is proposed to enhance the cross-correlation performance by incorporating multiple negative context images. 
As a multivariate operator, the SBCC is the closed-form solution of a well-designed optimization problem, which is formulated with both surrogate objective and correlation consistency objective. 
%
To our surprising, this framework also provides an alternative surrogate view for a set of generalized cross-correlation methods (PC, RPC, 1-CSPC, SPOF, etc).
On the correlation response, the SBCC method is more likely to achieve a desired Gaussian-shaped correlation response, encouraged by the objective function. 
And an arbitrary parameter $\mu(3.0)$ is fixed through parameter sensitivity analysis. 
Finally, the performance improvement of SBCC is verified with massive synthetic and real PIV image pairs.  
An interesting point is that  SBCC paves a new way for robust PIV cross correlation analysis via employing negative context images. 
Moving forward, we plan to apply SBCC beyond PIV to other tasks including one-dimensional time series, digital image correlation, acoustic imaging, etc.


%

\section*{Acknowledgment}
This work was supported by the National Natural Science Foundation of China (Grant No.: 52205575), Natural Science Foundation of Hubei Province (Grant Number: 2023AFB128) and Teaching Research Project of Wuhan University of Technology (Grant Number: W2022093). The authors would like to thank Dr. B. Wieneke and Zhenghao Cen for beneficial discussion.

\section*{Data Availability Statement}
The data that support the findings of this study are available from the corresponding author upon reasonable request.

\nocite{*}
\bibliography{ref}

\end{document}